\documentclass[sigconf]{acmart}
\usepackage{multirow}
\usepackage{listings}
\usepackage{rotating}

\usepackage{xcolor}
\definecolor{codegreen}{rgb}{0,0.6,0}
\definecolor{codegray}{rgb}{0.5,0.5,0.5}
\definecolor{codepurple}{rgb}{0.58,0,0.82}
\definecolor{backcolour}{rgb}{0.95,0.95,0.92}
\definecolor{zhen}{RGB}{0, 143, 67}

\lstdefinestyle{mystyle}{
  backgroundcolor=\color{backcolour}, commentstyle=\color{codegreen},
  keywordstyle=\color{magenta},
  numberstyle=\tiny\color{codegray},
  stringstyle=\color{codepurple},
  basicstyle=\ttfamily\footnotesize,
  breakatwhitespace=false,         
  breaklines=true,                 
  captionpos=b,                    
  keepspaces=true,                 
  numbers=left,                    
  numbersep=5pt,                  
  showspaces=false,                
  showstringspaces=false,
  showtabs=false,                  
  tabsize=2
}

\AtBeginDocument{%
  \providecommand\BibTeX{{%
    \normalfont B\kern-0.5em{\scshape i\kern-0.25em b}\kern-0.8em\TeX}}}

\copyrightyear{2023}
\acmYear{2023}
\setcopyright{rightsretained}
\acmConference[SC-W 2023]{Workshops of The International Conference on High Performance Computing, Network, Storage, and Analysis}{November 12--17, 2023}{Denver, CO, USA}
\acmBooktitle{Workshops of The International Conference on High Performance Computing, Network, Storage, and Analysis (SC-W 2023), November 12--17, 2023, Denver, CO, USA}
\acmDOI{10.1145/3624062.3624172}
\acmISBN{979-8-4007-0785-8/23/11}

\begin{document}

\newcommand{\aliasAPP}{HPC-GPT}
\title{\aliasAPP: Integrating Large Language Model for High-Performance Computing}


\author{Xianzhong Ding} 
\email{xding5@ucmerced.edu}
\affiliation{%
  \institution{University of California, Merced}
  \city{Merced}
  \state{CA}
  \country{USA}
  }
\affiliation{%
  \institution{Argonne National Laboratory }
  \city{Lemont}
  \state{IL}
  \country{USA}
  }

\author{Le Chen}
\email{lechen@iastate.edu}
\affiliation{%
  \institution{Iowa State University}
  \city{Ames}
  \state{IA}
  \country{USA}
}
\affiliation{%
  \institution{Lawrence Livermore National Laboratory}
  \city{Livermore}
  \state{CA}
  \country{USA}
  }

\author{Murali Emani} 
\email{memani@anl.gov}
\affiliation{%
  \institution{Argonne National Laboratory }
  \city{Lemont}
  \state{IL}
  \country{USA}
  }

\author{Chunhua Liao, Pei-Hung Lin, \\Tristan Vanderbruggen} 
\email{{liao6, lin32, vanderbrugge1}@llnl.gov}
\affiliation{%
  \institution{Lawrence Livermore National Laboratory}
  \city{Livermore}
  \state{CA}
  \country{USA}
  }

\author{Zhen Xie} 
\email{zxie3@binghamton.edu}
\affiliation{%
  \institution{Binghamton University}
  \city{Binghamton}
  \state{NY}
  \country{USA}
  }

\author{Alberto E. Cerpa, Wan Du} 
\email{{acerpa,wdu3}@ucmerced.edu}

\affiliation{%
  \institution{University of Calif  ornia, Merced}
  \city{Merced}
  \state{CA}
  \country{USA}
  }

\begin{abstract}
Large Language Models (LLMs), including the LLaMA model, have exhibited their efficacy across various general-domain natural language processing (NLP) tasks. However, their performance in high-performance computing (HPC) domain tasks has been less than optimal due to the specialized expertise required to interpret the model's responses. In response to this challenge, we propose \aliasAPP, a novel LLaMA-based model that has been supervised fine-tuning using generated QA (Question-Answer) instances for the HPC domain. To evaluate its effectiveness, we concentrate on two HPC tasks: managing AI models and datasets for HPC, and data race detection. By employing \aliasAPP, we demonstrate comparable performance with existing methods on both tasks, exemplifying its excellence in HPC-related scenarios. Our experiments on open-source benchmarks yield extensive results, underscoring \aliasAPP's potential to bridge the performance gap between LLMs and HPC-specific tasks. With \aliasAPP, we aim to pave the way for LLMs to excel in HPC domains, simplifying the utilization of language models in complex computing applications.

\end{abstract}

\ccsdesc[500]{Software and its engineering}
\ccsdesc[500]{Computing methodologies~Artificial intelligence}
\ccsdesc[500]{Computing methodologies~Parallel algorithms}

%
\keywords{High-performance Computing, Data Race Detection, Large Language Model, OpenMP, Neural Network.}

\maketitle

\section{Introduction}


Deep learning has found application across various domains and has been utilized in a multitude of applications, including code generation \cite{li2022competition, lachaux2020unsupervised}, building control \cite{ding2019octopus, ding2020mb2c}, irrigation scheduling \cite{ding2022drlic, ding2022smart}, and even in tasks related to epistemic uncertainty and occupancy estimation \cite{CLUE, TODOS}. These applications showcase the versatility and adaptability of deep learning techniques. On the other hand, language models (LMs), particularly large language models (LLMs) trained on extensive textual data, have recently demonstrated remarkable capabilities in various natural language processing and visualization tasks, reflecting their growing importance in the field of artificial intelligence. They have also been widely used to process programming languages due to the similarities between natural languages and programming languages. Based on an LLM trained on code~\cite{chen2021evaluating}, GitHub provides an AI assistant for developing software. 

With the surging popularity of LLMs, the high-performance computing (HPC) community is exploring their potential to tackle various HPC challenges like programming language processing, parallel programming, and question answering. However, existing pre-trained LLMs were initially designed for general tasks, such as dialogue and article summarization, which limits their effectiveness in HPC applications due to a lack of relevant domain knowledge. Several attempts have been made to employ LLMs for HPC tasks. For instance, Godoy et al. \cite{godoy2023evaluation} evaluated the capacity of OpenAI Codex \cite{Codex} via Copilot for generating HPC numerical kernels, while Chen et al. \cite{chen2023lm4hpc} developed pipelines to support common HPC tasks like code similarity analysis and parallelism detection. However, they are still in the process of investigating the adequacy of existing LLMs for HPC tasks. The general LLMs perform poorly on specific HPC tasks. For example, in question-answering tasks related to HPC, such as OpenMP Q$\&$A \cite{chen2023lm4hpc}, even the latest ChatGPT fails to provide accurate answers. This is because general LLMs lack the specific knowledge required to solve HPC-related challenges, despite being trained on large datasets that may include HPC-related information. Although the generalization ability of LLMs improves with more diverse training data from various domains, their performance in a specific domain, like HPC, reduces.

In this paper, we aim to explore and train an open-source LLM tailored for HPC applications. Our focus revolves around two general applications, each comprising two subtasks. The first application focuses on managing AI models and datasets for programming language processing (PLP) \cite{flynn2022finding} and MLPerf \cite{reddi2020mlperf} tasks. Through datasets and model information collection, we aim to help users better comprehend the required computing resources for training AI models and datasets. Liao et al. \cite{liao2021hpc} proposed HPC ontology to capture HPC-related concepts using the web ontology language (OWL) \cite{motik2009owl}. HPC ontology allows users to query PLP and HPC information but is limited by the manual processes of data creation, collection, and updates. Moreover, HPC ontology depends on users to articulate their requirements, including conditions or constraints related to code, hardware, dataset, models, and more, resulting in a challenge to accurately capture and utilize such nuanced semantics.

In the second application, we evaluate data race detection for C/C++ and Fortran tasks using the public data race benchmark (DRB) \cite{lin2021high}. Data race detection analyses can be broadly categorized into dynamic and static approaches. While dynamic approaches, like Intel Inspector \cite{intel_inspector}, are commonly used, there is growing interest in static data race detection tools that complement dynamic approaches. LLOV \cite{bora2020llov} is a recent tool evaluated with DRB. However, designing such tools demands significant data race expertise and running code snippets, making the process time-consuming.


Leveraging LLMs for solving HPC tasks holds tremendous promise, as a single HPC-domain LLM can effectively handle various general applications, eliminating the need to devise separate methods for each task. However, training an LLM for the HPC domain \cite{liu2022lobster, liu2022large} presents non-trivial challenges, primarily revolving around the collection of domain-specific training data. In contrast to traditional language model training, which involves feeding large volumes of raw data like text, fine-tuning an LLM for HPC necessitates specialized instruction data (e.g., PromptInstructions \cite{bach2022promptsource} and Super-Natural-Instructions \cite{yang2020generative}). These instructions guide the model's understanding of HPC concepts and tasks, ensuring it performs well in this domain. However, acquiring such instruction data proves to be a daunting task due to the associated costs and limitations in diversity. To address this issue, Wang et al. \cite{wang2022self} propose a semi-automated process for instruction-tuning a pre-trained language model using instructional signals extracted from the model itself. This approach mitigates some of the challenges in data collection and provides an alternative means of training the LLM for HPC tasks. However, it requires a set of manually-written tasks to guide the overall generation process, which can be time-consuming and resource-intensive.

\begin{figure*}[t]
\includegraphics[width=\textwidth]{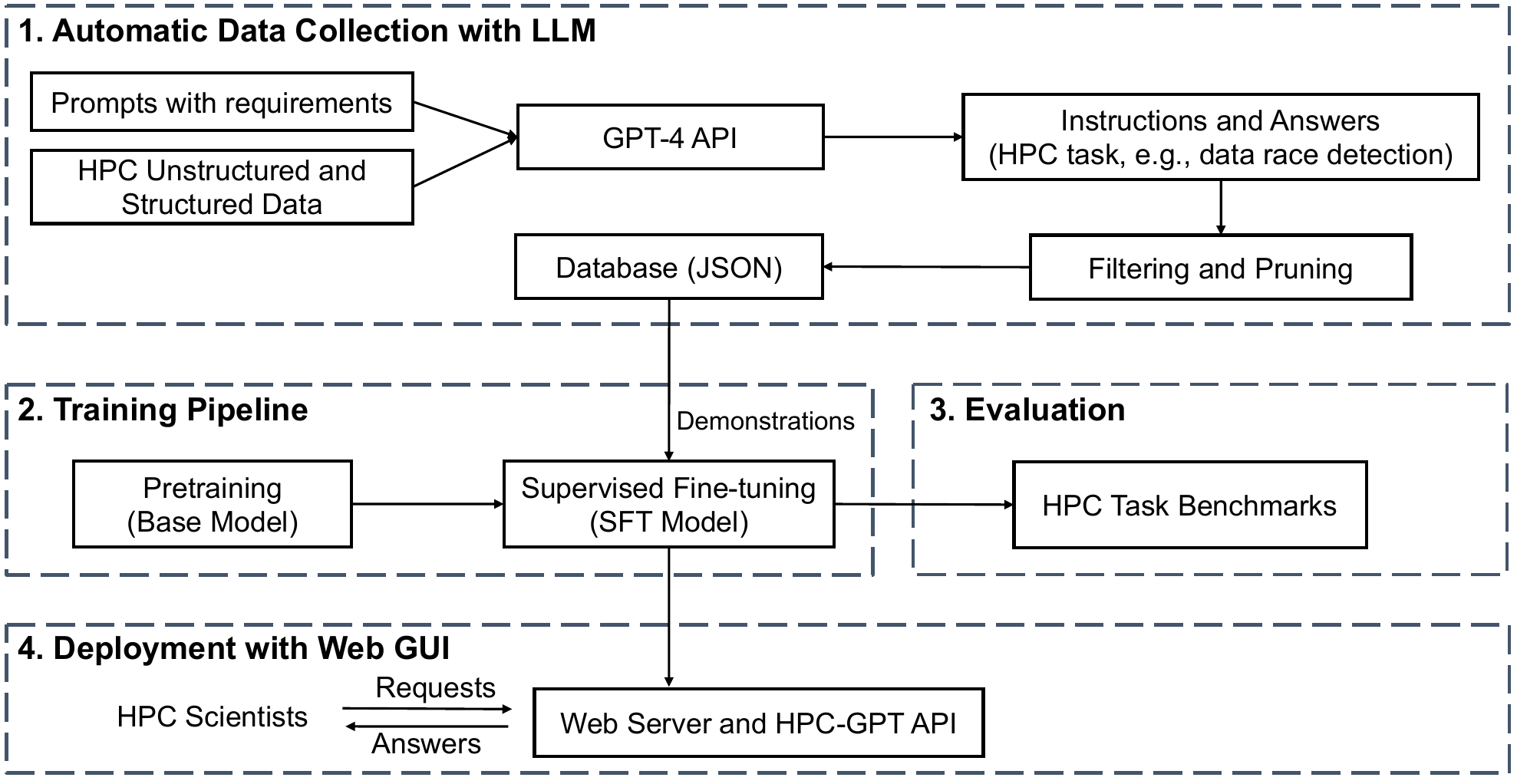}\Description{framework}
  \caption{HPC-GPT Architecture.}
  \label{framework}
\end{figure*}

In this paper, we present HPC-GPT, a specialized LLM tailored for the HPC domain. To facilitate fine-tuning, we design an automatic collection method to gather instruction data for training the open-source base LLaMa \cite{touvron2023llama} and LLaMa2 model \cite{touvron2023llama2}. Leveraging the power of the existing LLM (GPT-4), we generate domain-specific and contextually relevant instructions from unsupervised text. This approach ensures that the generated instructions are specific to the HPC domain and aligned with the provided unsupervised text. To address specific HPC tasks, we design distinct prompts that outline relevant requirements, enabling the model to generate instructions and answers separately. Subsequently, we establish filtering and pruning rules to eliminate instruction data that fails to meet the specified requirements. Our HPC-GPT model is built upon the foundation of the open-source LLaMa-13B base model, further enhancing its language generation capabilities. To showcase the potential of HPC-GPT, we utilize the two HPC applications mentioned earlier as illustrative examples to evaluate the performance. We summarize the main contributions of this paper as follows:
\begin{itemize}

  \item Introduction of the HPC-GPT model, representing the first open-source LLM fine-tuned using HPC instruction data. This specialized model is specifically designed to excel in HPC tasks.
  \item Integration of HPC knowledge into the model, ensuring it possesses accurate and domain-specific information. By incorporating HPC knowledge, the model leads to enhanced performance in HPC applications.
  \item Release of two open-source datasets for the HPC domain: one containing HPC models and datasets, and the other dedicated to data race detection. These datasets provide valuable resources for researchers in the HPC community, facilitating further advancements in the field. The code \footnote{https://github.com/dingxianzhong/HPC-GPT} and datasets \footnote{https://huggingface.co/datasets/HPC-GPT/HPC} are publicly available.
  \item Comparative evaluation against state-of-the-art methods. This progress underscores the model's effectiveness and its potential to outperform existing approaches in HPC-related tasks.

\end{itemize}

\section{Related Work}
\subsection{Large Language Models}

In the realm of language processing, recent strides in LLMs \cite{ChatGPT, bubeck2023sparks, chowdhery2022palm, ding2023hpcgpt} have showcased their superiority over earlier paradigms like pretraining \cite{lample2019cross} and fine-tuning \cite{howard2018universal}. This remarkable progress can be attributed to the substantial growth in model scale, resulting in qualitative shifts within LLMs, known as emergent abilities. These newfound capabilities encompass in-context learning, enabling the model to tackle zero-shot tasks, and the ability to follow chains of thought \cite{wei2022chain}, thereby enhancing its performance on intricate tasks. 

OpenAI's groundbreaking work in developing ChatGPT \cite{ChatGPT} and GPT-4~\cite{bubeck2023sparks} has brought about a paradigm shift in the understanding of LLMs. These models have demonstrated remarkable performance, yet OpenAI has maintained confidentiality regarding their training strategies and weight parameters. To address this limitation, LLaMa \cite{touvron2023llama} and LLaMa2 \cite{touvron2023llama2} emerge as an open-source alternative to GPT, available in sizes ranging from 7 billion to 65 billion parameters. While LLaMa's performance is comparable to GPT-3.5 in general tasks, it falls short in HPC-related tasks due to its training data being predominantly focused on general applications. 



\subsection{Large Language Models for HPC}
LLMs, such as GPT-4 and LLaMA, have been widely used in multiple domains, including natural language processing, visualization, and so on. However, applying them for analyzing and optimizing HPC tasks is still challenging due to the lack of HPC-specific support. To address this challenge, LM4HPC \cite{chen2023lm4hpc} represents the first attempt to adapt LLMs to the HPC domain. This is achieved by creating the LM4HPC framework to facilitate research into HPC analyses and optimizations using LMs. Tailored
for supporting HPC datasets, AI models, and pipelines, the LM4HPC framework is built on top of a range of components from different levels of the machine learning software stack, with Hugging Face-compatible APIs. However, the framework currently is still relying on existing general LLM for HPC tasks which is suboptimal.


\section{Design of \aliasAPP}
In this section, we describe in detail the design of HPC-GPT.
\subsection{HPC-GPT Overview}

Figure \ref{framework} illustrates an overview of the proposed \aliasAPP \\ framework, which can be divided into four main stages: HPC domain data collection, training, evaluation, and deployment. In the HPC domain data collection stage, we develop a data collection method that automatically gathers the necessary training data for two specific HPC applications. This data curation process ensures that the model is trained on relevant and domain-specific information. Moving on to the training stage, we employ supervised fine-tuning on the open-source LLM using the generated instruction data. Fine-tuning helps the model adapt to the intricacies of HPC-related tasks and enhances its performance in the targeted domain. In the evaluation stage, we rigorously assess the well-trained HPC-GPT model's capabilities on a public data race detection dataset. This evaluation process validates the model's effectiveness and performance in real-world scenarios. Finally, after successful training and evaluation, we deploy the HPC-GPT model to a web server, making it readily available for HPC scientists and researchers to utilize in their work.


\subsection{Automatic Data Collection with LLM}\label{sec:data collection}
\begin{lstlisting}[language=Python, caption=Instruction Generation Prompt, label=instructions]
"The HPC knowledge is: 

{unsupervised knowledge data}

According to the information above, please help me generate {number} questions. 

Here are the requirements:
1. Try not to repeat the verb for each question to maximize diversity.
2. Make sure the output is less than 50 words.
3. The questions can be asked under many conditions.
4. Do not generate the same or similar questions as generated before.

Now, please generate the instructions following the above requirements."
\end{lstlisting}
\vspace{-0.1in}

\textbf{Instruction Generation:} In this stage, we leverage the capabilities of the existing LLM, GPT-4, to generate domain-specific instructions based on unsupervised text, encompassing both unstructured and structured HPC data. The unsupervised text serves as the input and includes a wide range of HPC-related knowledge data. To tailor the instructions for specific HPC tasks, we design instruction prompts with relevant requirements, as depicted in Listing \ref{instructions}. These prompts guide GPT-4 to respond with instructions that align with the provided text data. The generated instructions serve as guidance for the subsequent stages of the process.

\begin{figure}[t]
\includegraphics[width=\columnwidth]{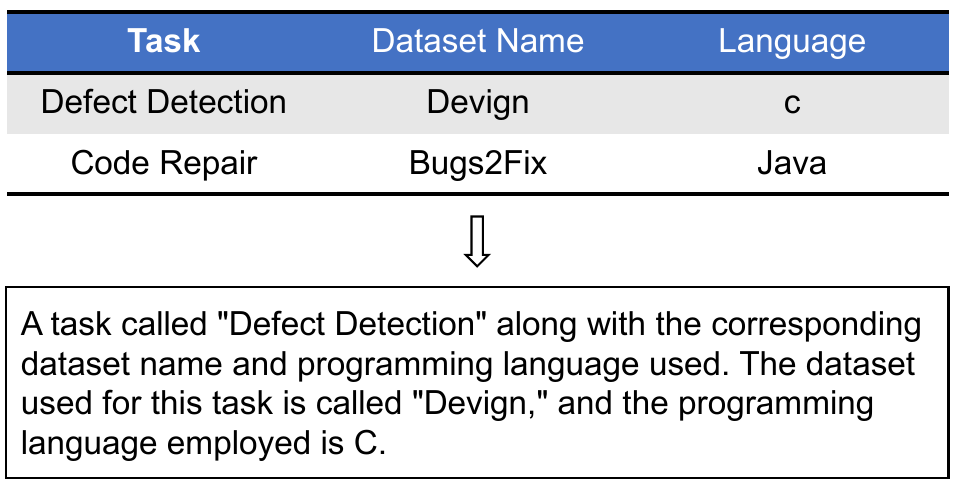}
  \caption{Transformation of unsupervised
structured data.}
  \label{fig: table_to_txt}
\end{figure}

The \textbf{unsupervised knowledge data} within the prompt consists of sequential text. Unstructured knowledge data, such as websites and conference papers, can be directly used after undergoing necessary cleaning processes. On the other hand, structured data, like tables, needs to be converted into unstructured textual data before it becomes usable. As illustrated in Figure \ref{fig: table_to_txt}, this conversion can be achieved through slot-filling using templates or by concatenating each data entry with its corresponding attribute name. These steps ensure that structured data is transformed into a format that can be effectively utilized by the language model for instruction generation and fine-tuning.



\textbf{Answer Generation:} In this stage, the language model is tasked with generating answers to the instruction questions based on the corresponding unsupervised HPC knowledge. The process can be expressed as $A = P(K, Q)$, where $K$, $Q$, and $A$ represent unsupervised knowledge, instruction question, and answer, respectively. Similar to the previous stage, the prompt for generating answers is provided, as shown in Listing \ref{answer}. Using the instruction questions and the associated unsupervised knowledge, the language model is prompted to produce accurate and contextually relevant answers. This step enhances the model's understanding of the provided knowledge and strengthens its ability to respond effectively to the instruction questions.

\begin{lstlisting}[language=C, caption=Instruction-Answer Generation Prompt, label=answer]
"The HPC knowledge is: 

{unsupervised knowledge data}.

Please answer the following question based on the above knowledge:
{the generated instruction}

Here are the requirements:
1. Try not to repeat the verb for each answer to maximize diversity.
2. Make sure the output is less than 50 words.
3. The questions can be asked under many conditions.
4. Make sure the answer is more than 10 words.
5. Make sure the answer can be obtained from the information provided. 
6. Do not generate the same or similar answers as generated before.
7. There are three fields for your generation: {"instruction": <question>, "Input":"", "output": <answer> }.

Now, please generate the data in JSON format following the above requirements."

\end{lstlisting}

\textbf{Filtering and Pruning:}
Although we explicitly instruct the model with the prompt in Listing \ref{answer} to not generate the same or similar instructions and answers as generated before, we still observe that the model produces instruction data that violates these rules. Additionally, the generated instances of instructions also exhibit cases where they do not adhere to the required format and become unparseable. Therefore, it is necessary to further filter out these problematic examples. To mitigate these issues, we implement a postprocessing step to filter out inappropriate responses and correct any formatting errors. This involves developing heuristics and rule-based methods to identify and remove instances that violate the instructed constraints. By applying these filters, we ensure that the generated text adheres to the predefined guidelines and maintains the desired level of correctness.








\subsection{Training Pipeline}
There are two stages in the training pipeline: pretraining and supervised fine-tuning.
\subsection{Pretraining}
When training LLM for HPC domain, the base model provides foundational language understanding, enabling quicker adaptation to the new domain's nuances. It offers transferable linguistic knowledge and reduces data requirements. The base model's broad comprehension aids in fine-tuning while retaining the ability to generate accurate responses beyond the specific domain. We select LLaMA  \cite{touvron2023llama} and LLaMA 2 \cite{touvron2023llama} as base models for pretraining. The reason is two-fold. Firstly, the LLaMA series are open-source and free to conduct research, in contrast to the commercial nature of the GPT series. Secondly, LLaMA outperforms other models, like GPT-3, in efficiency and resource utilization. LLaMA series is a collection of multi-lingual base models with parameters ranging from 7 billion to 70 billion. Here, We adopt the 13B version for both LLaMA and LLaMA 2 models, ensuring training accessibility and achieving superior performance.

\subsection{Supervised Fine-tuning}
Although LLMs exhibit remarkable performance in general domains, their
lack of domain-specific knowledge results in suboptimal performance in HPC fields that require specialized expertise. The HPC field’s inherent nature necessitates models to possess comprehensive knowledge bases for relevant queries. Supervised fine-tuning has proven to be effective in tuning LLM for different tasks \cite{wei2021finetuned, ouyang2022training}. Supervised fine-tuning helps the models perform satisfactorily under zero-shot scenarios with the cost of sufficient annotated instructions. Inspired by the automatic construction of the instruction along with the instances (inputs and outputs) \cite{wang2022self}, we generate our instruction data based on the above HPC knowledge with the proposed method in Section \ref{sec:data collection}. Table \ref{tab:instance examples} shows the supervised fine-tuning data examples for 2 HPC tasks. The data instance has an instruction that describes the tasks in natural language. In HPC-based tasks, we consider the instructions and input are the same and leave the input null value.



\begin{table*}[]
\caption{Instance with An Instruction}
\label{tab:instance examples}
\begin{tabular}{|l|}
\hline
\multicolumn{1}{|c|}{{\color[HTML]{000000} Task1: Model and datasets for HPC}}                                                                                                                                      \\ \hline
{\color[HTML]{000000} \begin{tabular}[c]{@{}l@{}}"instruction": "What kind of dataset can be used if the language is C/C++ and the baseline is CodeBERT?", \\ "input":"", \\ "output": "The POJ-104 dataset can be used for clone detection tasks if the language is C/C++ and the baseline is CodeBERT." \\ \end{tabular}}                                                                                                                                                                                                  \\ \hline
\multicolumn{1}{|c|}{Task 2: Data Race Detection}                                                                                                                                                                                                                                                                                                                                                                                                  \\ \hline
\begin{tabular}[c]{@{}l@{}}"instruction": "Given the code snippet: ```c\textbackslash{}n\#pragma omp parallel for\textbackslash{}nfor (i = 1; i \textless n; i++)\textbackslash{}n    y{[}i{]} = x{[}i{]} + y{[}i - 1{]};\textbackslash{}n```, \\  \hspace{1cm} \hspace{1cm}  help me detect if adding pragma will cause a data race problem? Answer 'yes' if it causes a data race problem\\ \hspace{1cm} \hspace{1cm}  and 'no' if it will not cause a data race problem.", \\ "input":"", \\ "output": "yes" \end{tabular} \\ \hline
\end{tabular}
\end{table*}



\section{Experiment}
\subsection{Experiments Setting} 

Our experimental setup takes place on a DGX A100 node within the ThetaGPU server, equipped with eight NVIDIA A100 Tensor Core GPUs and two AMD Rome CPUs, providing 320 GB GPU memory. For training \aliasAPP, we allocate 12 hours with 200 training epochs and set the learning rate to 2e-5. To optimize the process, we utilize a training and evaluation batch size of 16. To efficiently fine-tune \aliasAPP, we leverage fp16 precision to reduce memory requirements. Furthermore, we apply the LoRA \cite{hu2021lora} technique, known as low-rank adaptation, to reduce the number of trainable parameters. Additionally, we employ the PEFT \cite{liao2023parameter} technique, which stands for parameter-efficient fine-tuning, to further enhance the performance of the pre-trained models. 

\subsection{HPC Unstructured and Structured Data}
We collect the HPC raw data from GitHub, papers, and websites. For the PLP task, the unstructured data is from more than 40 papers related to PLP tasks, such as paper \cite{flynn2022finding}, and the structured data is from tables in the paper (e.g., \cite{lu2021codexglue}) and GitHub (e.g., \cite{CodeXGLUE}). For the MLPerf task, the unstructured data is from papers related to ML performance such as paper \cite{reddi2020mlperf}, and the structured data is from tables such as the results in the website \cite{mlperf_result}. We feed the raw data to LLM using the methods in Section \ref{sec:data collection} to collect the instruction data.

\subsection{Two HPC Tasks} 

\textbf{1. Managing AI Models and Datasets} has two subtasks: 1) PLP task and 2) MLPerf task. PLP task refers to leveraging machine learning techniques to understand and analyze programming languages in a human-readable manner. A large number of different ML models and datasets are published each year related to different kinds of PLP tasks such as code generation \cite{li2022competition}, clone detection \cite{li2022competition}, source-to-source translation \cite{lachaux2020unsupervised}, defect correction \cite{bovzivc2021mixed}, code documentation \cite{husain2019codesearchnet} and so on.
It is difficult for people, especially newcomers, to identify representative ones to get started. Different model architectures are used to solve different types of PLP tasks, ranging from program understanding to code generation. It is a daunting job for people to pick the right architectures for a given task. The goal of this task is to facilitate the reuse of datasets and models related to PLP tasks, so researchers or developers can easily create customized ML pipelines to solve a given task. MLPerf \cite{reddi2020mlperf} is a standardized benchmark designed to evaluate and compare the inference performance of machine learning models and frameworks. The benchmark includes different types of models and uses cases. Participants in the MLPerf benchmark must run their inference engines on a predefined set of models and datasets, following specified guidelines for hardware and software configurations. The goal of MLPerf task is to help the participants know the current exact system (e.g., 16-nodes-SPR-pytorch), processor (e.g., Intel(R) Xeon(R) Platinum 8462Y+), accelerator (e.g, NVIDIA H100-SXM5-80GB), and software (e.g., PyTorch NVIDIA Release 23.04) that can facilitate them to build similar ML models and datasets efficiently.

\textbf{2. Data Race Detection} is a critical task in HPC. It is aimed at finding data race bugs in multithreaded programs such as those using OpenMP. 
In general, a data race occurs when two or more threads perform conflicting accesses (with at least one access being a write) to a shared variable without any synchronization among the threads.
In this task, we cover two programming languages (such as C/C++ and Fortran) using OpenMP. The goal of this task is to detect if there is a data race problem given the code snippet.

\subsection{Baselines}
\aliasAPP~ comprises two versions: HPC-GPT (L1) and HPC-GPT (L2), each integrating the respective base models, LLaMa and LLaMa 2. In order to demonstrate the superior performance of HPC-GPT, we conduct a comparative analysis with different baseline methods. 

\textbf{Task 1: Managing AI Models and Datasets}

\begin{itemize}
  \item \textbf{ChatGPT} is the state of the art of commercial large language model, we use GPT-4 as its model engine with superior performance.

  \item \textbf{HPC Ontology \cite{liao2021hpc}} is a unified ontology for managing AI models and datasets for HPC.

\end{itemize}

\textbf{Task 2: Data Race Detection}

\begin{itemize}
  \item \textbf{LLOV \cite{bora2020llov}} is a fast, lightweight, language agnostic, and static data race checker for OpenMP programs based on the LLVM compiler framework.
  \item \textbf{Intel Inspector \cite{intel_inspector}} is a dynamic analysis
tool that detects threading and memory errors in C, C++ and Fortran
codes. It supersedes Intel’s Thread Checker tool, with added memory error checking. Supported thread errors include race conditions and deadlocks.
  \item \textbf{ROMP \cite{gu2018dynamic}} is a tool for detecting data races in executions of scalable parallel applications that employ OpenMP for node-level parallelism. 
  \item \textbf{Thread Sanitizer \cite{serebryany2009threadsanitizer}} is a runtime
data race detector developed by Google. ThreadSanitizer is now part of the LLVM and GCC compilers to enable data race detection for C++, Fortran and Go code.

  \item \textbf{GPT-3.5 and GPT-4} are standard commercial LLM developed by OpenAI. GPT-4 with 1.7 trillion parameters is the latest and most advanced version of GPT compared with GPT-3 with 175 billion parameters.

  \item \textbf{LLaMA \cite{touvron2023llama} and LLaMA 2 \cite{touvron2023llama2}} are standard open-source LLM developed by Meta. LLaMA 2 was trained on 40\% more data than LLaMA. We use 13B versions for LLaMA and LLaMA 2 due to the tradeoff between performance and training costs.

\end{itemize}


\subsection{Metrics} \label{sec: metrics}
To evaluate \aliasAPP, we calculate counts of True Positive (TP), False Positive (FP), True Negative (TN), and False Negatives (FN) based on tool results and ground truth. We use five standard metrics: Recall, Specificity, Precision, Accuracy, and F1 score to evaluate the quality of tools. We also report tool support rate (TSR), which is the ratio of how many test files are supported by a tool. The F1 score is a measure combining both precision and recall. It provides a single metric that weights precision and recall in a
balanced way, requiring both to have higher values for the F1-score value to rise. The reported adjusted F1 score, F1 score multiplied by the TSR, can show the true ability of a tool.



\begin{table}[]
\small 
\caption{Dataset Information for Task 1}
\label{tab:TASK1}
\begin{tabular}{|c|c|c|c|}
\hline
Subtasks      & Category                 & Number & Percentage \\ \hline
\multirow{13}{*}{PLP}   & Performance Modeling     & 44     & 7.30\%     \\ \cline{2-4} 
                        & Algorithm Classification & 41     & 6.80\%     \\ \cline{2-4} 
                        & Defect detection         & 47     & 7.79\%     \\ \cline{2-4} 
                        & Clone detection          & 45     & 7.46\%     \\ \cline{2-4} 
                        & Code Completion          & 39     & 6.47\%     \\ \cline{2-4} 
                        & Compiler Analyses        & 37     & 6.14\%     \\ \cline{2-4} 
                        & Code Repair              & 48     & 7.96\%     \\ \cline{2-4} 
                        & Code Translation         & 41     & 6.80\%     \\ \cline{2-4} 
                        & Cloze Testing            & 48     & 7.96\%     \\ \cline{2-4} 
                        & Text-to-Code Generation  & 58     & 9.62\%     \\ \cline{2-4} 
                        & Code Summarization       & 48     & 7.96\%     \\ \cline{2-4} 
                        & Document Translation     & 52     & 8.62\%     \\ \cline{2-4} 
                        & Code Search              & 55     & 9.12\%     \\ \hline
\multirow{5}{*}{MLPerf} & Submitter                & 324    & 17.80\%    \\ \cline{2-4} 
                        & System                   & 386    & 21.21\%    \\ \cline{2-4} 
                        & Processor                & 347    & 19.07\%    \\ \cline{2-4} 
                        & Accelerator              & 362    & 19.89\%    \\ \cline{2-4} 
                        & Software                 & 401    & 22.03\%    \\ \hline
\end{tabular}
\end{table}

\begin{table*}[]
\caption{Dataset Information for Task 2}
\small 
\label{tab:task2}
\centering
\begin{tabular}{|c|c|cccccccccccccc|}
\hline
\multirow{3}{*}{Subtasks}  & \multirow{3}{*}{Metric} & \multicolumn{14}{c|}{Category}                                                                                                                                                                                                                                                                                                                                                                                                                                                                                                                                                                                                                                                                                                           \\ \cline{3-16} 
                         &                         & \multicolumn{7}{c|}{Code snippet with data races}                                                                                                                                                                                                                                                                                                 & \multicolumn{7}{c|}{Code snippet without data races}                                                                                                                                                                                                                                                                                                                                                                  \\ \cline{3-16} 
                         &                         & \multicolumn{1}{c|}{\begin{sideways}Unresolvable dependencies\end{sideways}} & \multicolumn{1}{c|}{\begin{sideways}Missing data sharing clauses\end{sideways}} & \multicolumn{1}{c|}{\begin{sideways}Missing synchronization\end{sideways}} & \multicolumn{1}{c|}{\begin{sideways}SIMD data races\end{sideways}} & \multicolumn{1}{c|}{\begin{sideways}Accelerator data races\end{sideways}} & \multicolumn{1}{c|}{\begin{sideways}Undefined behavior\end{sideways}} & \multicolumn{1}{c|}{\begin{sideways}Numerical kernel data races\end{sideways}} & \multicolumn{1}{c|}{\begin{sideways}Single thread execution\end{sideways}}

                        & \multicolumn{1}{c|}{\begin{sideways}Use of data sharing clauses\end{sideways}} & \multicolumn{1}{c|}{\begin{sideways}Use of synchronization\end{sideways}} & \multicolumn{1}{c|}{\begin{sideways}Use of SIMD directives\end{sideways}} & \multicolumn{1}{c|}{\begin{sideways}Use of accelerator directives\end{sideways}} & \multicolumn{1}{c|}{\begin{sideways}Use of special language features\end{sideways}} & \begin{sideways}Numerical kernels \end{sideways}\\ \hline
\multirow{2}{*}{C/C++}   & Number                  & \multicolumn{1}{c|}{132}                       & \multicolumn{1}{c|}{129}                          & \multicolumn{1}{c|}{130}                     & \multicolumn{1}{c|}{124}             & \multicolumn{1}{c|}{110}                    & \multicolumn{1}{c|}{128}                & \multicolumn{1}{c|}{133}                         & \multicolumn{1}{c|}{133}                                                                                           & \multicolumn{1}{c|}{105}                         & \multicolumn{1}{c|}{144}                    & \multicolumn{1}{c|}{119}                    & \multicolumn{1}{c|}{118}                           & \multicolumn{1}{c|}{126}                              & 131               \\ \cline{2-16} 
                         & Percentage              & \multicolumn{1}{c|}{7.49\%}                    & \multicolumn{1}{c|}{7.32\%}                       & \multicolumn{1}{c|}{7.38\%}                  & \multicolumn{1}{c|}{7.04\%}          & \multicolumn{1}{c|}{6.24\%}                 & \multicolumn{1}{c|}{7.26\%}             & \multicolumn{1}{c|}{7.55\%}                      & \multicolumn{1}{c|}{7.55\%}                                                                                        & \multicolumn{1}{c|}{5.96\%}                      & \multicolumn{1}{c|}{8.17\%}                 & \multicolumn{1}{c|}{6.75\%}                 & \multicolumn{1}{c|}{6.70\%}                        & \multicolumn{1}{c|}{7.15\%}                           & 7.43\%            \\ \hline
\multirow{2}{*}{Fortran} & Number                  & \multicolumn{1}{c|}{125}                          & \multicolumn{1}{c|}{103}                             & \multicolumn{1}{c|}{117}                        & \multicolumn{1}{c|}{122}                & \multicolumn{1}{c|}{101}                       & \multicolumn{1}{c|}{109}                   & \multicolumn{1}{c|}{111}                            & \multicolumn{1}{c|}{98}                                                                                              & \multicolumn{1}{c|}{126}                            & \multicolumn{1}{c|}{105}                       & \multicolumn{1}{c|}{130}                       & \multicolumn{1}{c|}{97}                              & \multicolumn{1}{c|}{108}                                 &       124            \\ \cline{2-16} 
                         & Percentage              & \multicolumn{1}{c|}{7.93\%}                          & \multicolumn{1}{c|}{6.54\%}                             & \multicolumn{1}{c|}{7.42\%}                        & \multicolumn{1}{c|}{7.74\%}                & \multicolumn{1}{c|}{6.41\%}                       & \multicolumn{1}{c|}{6.91\%}                   & \multicolumn{1}{c|}{7.04\%}                            & \multicolumn{1}{c|}{6.21\%}                                                                                              & \multicolumn{1}{c|}{8.00\%}                            & \multicolumn{1}{c|}{6.66\%}                       & \multicolumn{1}{c|}{8.25\%}                       & \multicolumn{1}{c|}{6.15\%}                              & \multicolumn{1}{c|}{6.85\%}                                 &      7.86\%             \\ \hline
\end{tabular}
\end{table*}

\subsection{Instruction Datasets for HPC}
We have collected a total of 5.86k instruction data for two HPC applications. The dataset details for task 1 and task 2 are presented in Table \ref{tab:TASK1} and Table \ref{tab:task2}, respectively. In Table \ref{tab:TASK1}, we focus on task 1: Managing AI models and datasets for PLP tasks and MLPerf tasks. There are 13 categories in PLP tasks covering various programming language processing subtasks, such as clone detection and code repair. The percentage range for these categories is 6.14\% to 9.62\%. Additionally, there are 5 categories dedicated to MLPerf tasks, such as the system and software, with a percentage range of 17.80\% to 22.03\%. In Table \ref{tab:task2}, the data race detection task has two main categories: "code snippet with data races" and "code snippet without data races." We have summarized 7 common data race types for both C/C++ and Fortran, including "missing data sharing" and "Unresolvable dependencies." Additionally, we have identified 7 types of data without data races, such as "Single thread execution" and "Use of synchronization." The percentage range for these categories is 5.96\% to 8.17\% for C/C++ and 6.15\% to 8.25\% for Fortran.

Overall, we have ensured that the number of different categories is balanced for various subtasks. This approach prevents one category from dominating and allows the LLM to maintain a well-rounded knowledge base for each task. This diverse and comprehensive dataset collection is crucial in fine-tuning the LLM effectively and preparing it to tackle a wide array of HPC-specific challenges with improved accuracy and performance.

\subsection{Preliminary Results}

\subsubsection{Managing AI models and Datasets}
This section presents some use cases using HPC-GPT for providing the dataset and model information and answering questions.

1) PLP Task. The AI models and dataset are fundamental information for PLP tasks \cite{chen2023learning, chen2023data, chen2022multi}. We use natural language to describe the code translation-related questions in \cite{flynn2022finding}.

\begin{lstlisting}[language=Python, caption=PLP Task Example, label=result_plptask]
Question: "What kind of dataset can be used for code translation tasks if the source language is Java and the target language is C#?"

Answer (GPT-4): "For code translation tasks from Java to C#, you would need a dataset that consists of pairs of Java code and their corresponding equivalent C# code. "

Answer (HPC-Ontology): "CodeTrans dataset"

Answer (HPC-GPT (L2)): "The CodeTrans dataset can be used for code translation tasks if the source language is Java and the target language is C#."
\end{lstlisting}

2) MLPerf Task. The system, processor, and software-related details are important for MLPerf tasks. We use natural language to describe the MLPerf-related questions in \cite{mlperf_result}.

\begin{lstlisting}[language=Python, caption=MLPerf Task Example, label=result_mlperftask]
Question: "What is the System if the Accelerator used is NVIDIA H100-SXM5-80GB and the Software used is MXNet NVIDIA Release 23.04?"

Answer (GPT-4): "As of my last update in September 2021, the NVIDIA H100-SXM5-80GB is a data center GPU designed for high-performance computing and deep learning workloads. "

Answer (HPC-Ontology): "dgxh100_n64"

Answer (HPC-GPT (L2)): "If the Accelerator used is NVIDIA H100-SXM5-80GB and the Software used is MXNet NVIDIA Release 23.04, the System is dgxh100_n64."

\end{lstlisting}

\begin{table}[]
\caption{Data Race Detection Tool and Compiler Version}
\label{tab:tool version}
\begin{tabular}{|c|c|c|}
\hline
Tools            & Version               & Compiler                \\ \hline
ThreadSanitizer & 10.0.0                & Clang/LLVM 10.0.0       \\ \hline
Intel Inspector & 2021.1  & Intel Compiler 2021.3.0 \\ \hline
ROMP            & 20ac93c               & GCC/gfortran 7.4.0      \\ \hline
LLOV            & N/A                   & Clang/LLVM 6.0.1        \\ \hline
\end{tabular}
\end{table}


\begin{table*}[]
\caption{Results of Individual Data Race Detection Tools and LLM-Based methods.}
\label{tab:data_race_result}
\begin{tabular}{|c|c|c|c|c|c|c|c|c|c|c|c|}
\hline
Tool             & Language                 & TP          & FP         & TN          & FN          & Recall          & Specificity     & Precision       & Accuracy        & TSR             & Adjusted F1     \\ \hline \hline
LLOV           & \multirow{10}{*}{C/C++}   & 58          & \textbf{9} & 78          & 29          & 0.6666          & 0.8965          & 0.8656          & 0.7816          & 0.9613         & 0.7532          \\ \cline{1-1} \cline{3-12} 
Intel Inspector  &                          & \textbf{76} & 41         & 46          & \textbf{10}  & \textbf{0.837} & 0.5287          & 0.6495          & 0.7052          & 0.9558          & 0.7487          \\ \cline{1-1} \cline{3-12} 
ROMP             &                          & 63          & 12         & 65          & 18          & 0.775          & 0.8333          & 0.8266          & 0.8037          & 0.8729    & 0.8000         \\ \cline{1-1} \cline{3-12} 
Thread Sanitizer &                          & 69          & \textbf{1} & \textbf{89} & 20          & 0.7752          & \textbf{0.9888} & \textbf{0.9857} & \textbf{0.8826} & \textbf{0.9889}          & \textbf{0.8679} \\ \cline{1-1} \cline{3-12} 
GPT-3.5            &                        &       52         &   36         &    45         &  30           &     0.6341           & 0.5555                &     0.5909          &    0.5951          &   0.9209              &     0.6117          \\ 
 \cline{1-1} \cline{3-12}
GPT-4            &                        &       65         &   31         &    50         &  17           &     0.7926            &  0.6172                &     0.6770           &     0.7055          &   0.9209              &     0.73033          \\ \cline{1-1} \cline{3-12} 
\cline{1-1} \cline{3-12} 
LLaMa         &                         &   65         &  61         &   20           &    17       &     0.7926              &  0.2469               &   0.5158             &            0.52147     &      0.9209           &        0.625         \\ 
\cline{1-1} \cline{3-12} 
LLaMa2         &                         &   71          &  66         &   15         &    11        &     0.8658              &  0.1851               &  0.51824             &           0.5276      &      0.9209           &        0.6484         \\ \cline{1-1} \cline{3-12} 
HPC-GPT (L1)          &                         &     64         &       20     &      61       &    18         &      0.7804             &   0.7530            &      0.7619         &      0.7668          &   0.9209              &     0.7710  \\ \cline{1-1} \cline{3-12} 
HPC-GPT (L2)           &                         &     67         &       17     &      64       &    15         &      0.8171             &   0.7901              &      0.7976         &      0.8037           &   0.9209              &     0.8072            \\ \hline \hline
LLOV           & \multirow{10}{*}{Fortran} & 40          & 11 & 70          & 36          & 0.5263 & 0.8641    & 0.7843    & 0.7006 & 0.9457          & 0.6299          \\ \cline{1-1} \cline{3-12} 
Intel Inspector  &                          & 66 & 11          & 65 & 17          & 0.7951         & 0.8552          & 0.8571          & 0.8238          & 0.9464 & 0.825 \\ \cline{1-1} \cline{3-12} 
ROMP             &                          & 57          & 10         & 54          & 20          & 0.7402          & 0.8437          & 0.8507          & 0.7872          & 0.8392 & 0.7916          \\ \cline{1-1} \cline{3-12} 
Thread Sanitizer &                          & 52          & \textbf{0} & 65 & 15 & 0.7761          & \textbf{1.0}    & \textbf{1.0}    & \textbf{0.8863}             & 0.7857          & 0.8739          \\ \cline{1-1} \cline{3-12} 
GPT-3.5            &                          &         54    &         31   &     50        &   31          &     0.6352            &            0.6172     &       0.6352           &     0.6265            &       \textbf{1.0}          &    0.6352             \\ 
\cline{1-1} \cline{3-12} 
GPT-4            &                          &         67    &         37   &     44        &   18          &     0.7882            &            0.5432     &      0.6442           &     0.6687            &        \textbf{1.0}         &    0.7089             \\ 
\cline{1-1} \cline{3-12} 
LLaMa         &                         &   63         &  55         &   26           &    22       &    0.7411              &  0.3209              &   0.5338             &            0.5361     &      \textbf{1.0}           &        0.6206        \\ 
\cline{1-1} \cline{3-12} 
LLaMa2         &                         &   59          &  63         &   18         &    26        &    0.6941              &  0.2222               &   0.4836             &           0.4638      &      \textbf{1.0}           &        0.5700         \\

\cline{1-1} \cline{3-12} 
HPC-GPT (L1)          &                         &   66          &  20         &   61           &    19        &     0.7764              &   0.7530               &    0.7674            &            0.7650      &      \textbf{1.0}           &       0.7719        \\ 
\cline{1-1} \cline{3-12} 
HPC-GPT (L2)          &                         &  \textbf{70}         &  15         &   \textbf{68}          &    \textbf{13}        &     \textbf{0.8433}              &  0.8192               &   0.8235            &           0.8313    &     \textbf{1.0}            &          \textbf{0.8333}       \\ 

\hline

\end{tabular}
\end{table*}

\textbf{Results.} 
The results for the PLP task and MLPerf task are shown in Listing \ref{result_plptask} and \ref{result_mlperftask} respectively. Upon analyzing the outputs, it is evident that \aliasAPP~ demonstrates its superiority in handling these tasks compared to GPT-4. In the PLP task, when the user asks, "What kind of dataset can be used for code translation tasks if the source language is Java and the target language is C\#?", \aliasAPP~ is capable of providing an exact answer with the "CodeTrans dataset," while GPT-4 merely repeats the question due to its lack of relevant knowledge. Similarly, in the MLPerf task, when the user presents a system-related question such as, "What is the System if the Accelerator used is NVIDIA H100 - SXM5 - 80 GB and the Software used is MXNet NVIDIA Release 23.04?", \aliasAPP~ promptly offers a specific answer, "dgxh100\_n64," demonstrating its understanding of the question. Conversely, ChatGPT's response includes a general introduction of the Accelerator but fails to provide the correct answer, indicating its lack of relevant knowledge regarding the MLPerf task.

To address this limitation, HPC-Ontology can also deliver accurate answers by leveraging SPARQL query language. However, it requires manual effort to write SPARQL queries for different questions and answers, making it less scalable. \aliasAPP~ overcomes this challenge by effectively translating human language into embeddings, enabling it to process all the information and provide relevant answers to various queries. In both tasks, \aliasAPP~ demonstrates its versatility in handling different conditions and benefits users by allowing them to express their questions freely while receiving accurate and related answers.

\vspace{-0.15in}
\subsubsection{Data Race Detection}

In this section, we present the evaluation results of \aliasAPP~ on DataRaceBench V1.4.0 \cite{lin2021high, chen2023dataracebench} with 177 C/C++ test programs and 166 Fortran test programs. Among these, 88 C/C++ and 84 Fortran test cases exhibit data races, while 89 C/C++ and 82 Fortran test cases are free from data races. We utilize the metrics defined in Section \ref{sec: metrics} to assess the performance of data race detection tools (LLOV, Intel Inspector, ROMP, and Thread Sanitizer) and LLM-based methods (GPT-3.5, GPT-4, LLaMa, LLaMa2, HPC-GPT (L1) and HPC-GPT (L2)). The compiler version is shown in Table \ref{tab:tool version}. The results are summarized in Table \ref{tab:data_race_result}, where the best result for each metric is highlighted in bold. In C/C++ language, Thread Sanitizer excels in Adjusted F1 (0.8679), specificity (0.9888), precision (0.9857), and accuracy (0.8826). Remarkably, HPC-GPT (L2) secures second place in accuracy (0.8037), Adjusted F1 (0.8072), and Recall (0.8171). For Fortran, Thread Sanitizer shines with the best results in specificity (1.0), precision (1.0), and accuracy (0.8863). HPC-GPT (L2) leads in Recall (0.8433) and Adjusted F1 (0.8333).


Compared with LLM-based methods, both HPC-GPT (L1) and HPC-GPT (L2) demonstrate notable improvements over LLaMa, LLaMa 2, GPT-3.5, GPT-4 across five key metrics (recall, specificity, precision, accuracy, and adjusted F1). In C/C++ language, HPC-GPT (L2) achieves improvements of 36.11\%, 34.84\%, 26.33\%, 11.1\%, and 3.85\% compared with LLaMa, LLaMa 2, GPT-3.5, GPT-4, and HPC-GPT (L1). For Fortran, HPC-GPT (L2) attains enhancements of 31.89\%, 35.23\%, 21.34\%, 15.79\%, and 7.28\% over LLaMa, LLaMa 2, GPT-3.5, GPT-4, and HPC-GPT (L1). Notably, all LLM-based methods share the same TSR due to an 8k token constraint, limiting input length. For C/C++, TSR is lower than existing tools, with 14 test cases exceeding 8k tokens. Conversely, Fortran's TSR for LLM-based methods is 1.0, surpassing existing tools.



\section{Challenges and Possible Solutions}

\textbf{The Token Length of Existing LLMs:} The limitation of LLM-based methods in detecting data races that exceed the 8k token limit poses a significant challenge in practical applications. When dealing with real-world software projects, data races may involve extensive sections of code that surpass the token limit. As a consequence, LLM-based methods fail to process such data race instances, leading to incomplete and potentially inaccurate results. To address this limitation, One approach is to investigate ways to extend the token limit of LLM models, allowing them to handle longer code snippets effectively. This involves optimizing the model architecture or leveraging advanced hardware configurations. Another avenue for improvement involves devising a pre-processing or partitioning mechanism to break down large code snippets into smaller, manageable segments that fit within the token limit. This way, LLM-based methods can analyze each segment individually and then combine the results to detect data races across the entire codebase.


\textbf{How to update \aliasAPP~ with Latest Data:}
Updating HPC-GPT with the latest data poses a challenge due to the continuous release of datasets and models. Several strategies can be employed for effective updates. Initially, new data can be periodically gathered to retrain the entire HPC-GPT alongside existing data. Another method involves creating a checkpoint of the current model version and then resuming training using the newly acquired data. Another approach leverages the LangChain framework \cite{LangChain}, wherein HPC-GPT integrates new data seamlessly. The Longchain APIs enable the storage of text within semantic vector stores. This integration process entails the division of text into chunks, followed by embedding and matching prompts with the most relevant vector chunks. Consequently, this enhances the context of responses while adhering to token limitations.

\section{Conclusion}

This paper proposes HPC-GPT, a large language model specifically designed for the HPC domain. We explore the capabilities of this model in addressing two common tasks in HPC and demonstrate its excellent performance. In the first task, concerning model and dataset selection for HPC, HPC-GPT exhibits the ability to retrieve and extract HPC-related datasets and models based on human expressions. It efficiently handles multiple conditions, providing valuable support for researchers in the HPC field. Additionally, we evaluate HPC-GPT's performance in data-race detection in an OpenMP program, a critical concern in parallel computing. Our experiments demonstrate that HPC-GPT achieves good performance compared to existing data-race detectors. 

\begin{acks}
This research was funded in part by and used resources at the
Argonne Leadership Computing Facility, which is a DOE
Office of Science User Facility supported under Contract
DE-AC02-06CH11357. This work is also 
prepared by LLNL under Contract DE-AC52-07NA27344 (LLNL-CONF-853156) and supported by the U.S. Department of Energy, Office of Science, Advanced Scientific Computing Research.
\end{acks}

\bibliographystyle{ACM-Reference-Format}
\bibliography{sample-base}

\appendix

\end{document}